\documentclass[useAMS]{mn2e}%,usenatbib
\usepackage{graphicx}
\usepackage{txfonts}
\usepackage{color}

\title[Unveiling the 3D temperature structure of galaxy clusters by means of the
thermal SZ effect]{Unveiling the 3D temperature structure of galaxy
clusters by means of the thermal Sunyaev-Zel'dovich effect}

\author[D. A. Prokhorov, Y. Dubois, S. Nagataki, T. Akahori, and K. Yoshikawa]{D. A. Prokhorov$^{1}$\thanks{E-mail:
phdmitry@stanford.edu}, Y, Dubois$^{2}$, S. Nagataki$^{3}$, T. Akahori$^4$, and K. Yoshikawa$^5$\\
$^{1}$Hansen Experimental Physics Laboratory, Stanford University,
Stanford, CA 94305, USA\\
$^{2}$Astrophysics, University of Oxford, Denys Wilkinson Building,
Keble Road, Oxford, OX13RH, United Kingdom\\
$^{3}$ Yukawa Institute for Theoretical Physics, Kyoto University,
Kitashirakawa Oiwake-cho, Sakyo-ku, Kyoto, 606-8502, Japan\\
$^{4}$ Research Institute of Basic Science, Chungnam National
University, Daejeon, Republic of Korea\\
$^{5}$ Center for Computational Sciences, University of Tsukuba,
1-1-1, Tennodai, Ibaraki 305-8577, Japan}
\begin{document}

%\date{Accepted 1988 December 15. Received 1988 December 14; in
%original form 1988 October 11}

\pagerange{\pageref{firstpage}--\pageref{lastpage}} \pubyear{2002}

\maketitle

\label{firstpage}

\begin{abstract}
The Sunyaev-Zel'dovich (hereafter SZ) effect is a promising tool to
derive the gas temperature of galaxy clusters. The approximation of
a spherically symmetric gas distribution is usually used to
determine the temperature structure of galaxy clusters, but this
approximation cannot properly describe merging galaxy clusters. The
methods used so far, which do not assume the spherically symmetric
distribution, permit us to derive 2D temperature maps of merging
galaxy clusters. In this paper, we propose a method to derive the
standard temperature deviation and temperature variance along the
line-of-sight, which permits us to analyze the 3D temperature
structure of galaxy clusters by means of the thermal SZ effect. We
also propose a method to reveal merger shock waves in galaxy
clusters by analyzing the presence of temperature inhomogeneities
along the line-of-sight.
\end{abstract}


\section{Introduction}

Galaxy clusters are large gravitationally bound structures of a size
of $\sim$1--3 Mpc, which have arisen in the hierarchical structure
formation of the Universe. Intergalactic gas fills the space between
galaxies in the clusters. The typical number density and temperature
values of this intergalactic gas are $10^{-1}$--$10^{-3}$ cm$^{-3}$
and 2-10 keV, respectively(for a review, see e.g., Sarazin 1986).
Inverse Compton scattering between cosmic microwave background (CMB)
photons and hot free electrons in clusters of galaxies causes a
change in the intensity of the CMB radiation towards clusters of
galaxies (the Sunyaev-Zel'dovich effect, hereinafter the SZ effect;
for a review, see Sunyaev \& Zel'dovich 1980).

The SZ effect is an important tool for the study of clusters of
galaxies (for a review, see Birkinshaw 1999), since both the
amplitude and spectral form of the CMB distortion caused by the SZ
effect depend on the intergalactic gas parameters (such as the gas
number density and temperature). A relativistically correct
formalism for the SZ effect based on the probability distribution of
the photon frequency shift after scattering was given by Wright
(1979) to describe the Comptonization process of soft photons by
high temperature plasma. SZ relativistic corrections are significant
even for galaxy clusters with temperatures of 3-5 keV and allow us
to derive the gas temperature from SZ observations (see, e.g.,
Colafrancesco \& Marchegiani 2010). The constraints on the gas
temperature of high temperature clusters from measurements of SZ
relativistic corrections have been firstly obtained by Pointecouteau
et al. (1998) and Hansen et al. (2002). The imaging and
spectroscopic capabilities of coming SZ experiments will permit us
to analyze the gas temperature structure even in relatively cool
galaxy clusters (Colafrancesco \& Marchegiani 2010).

Using hydrodynamic simulations of galaxy clusters, Kay et al. (2008)
compared the temperatures derived from the X-ray spectroscopy (the
X-ray temperature) and from the SZ effect (the SZ temperature). They
demonstrated that the SZ temperature corresponds to the
Compton-averaged electron temperature for relatively cool galaxy
clusters. Their SZ temperature map is directly calculated from the
simulated temperature and density distributions.

Colafrancesco \& Marchegiani (2010) show how to obtain detailed
information about the cluster temperature distribution for
spherically--symmetric galaxy clusters. However, the approximation
of a spherical symmetry cannot properly describe merging galaxy
clusters. Signatures of shock waves caused by galaxy cluster mergers
are imprinted on the 2D X-ray surface brightness maps of merging
galaxy clusters (for a review, see Markevitch \& Vikhlinin 2007)
and, therefore, merger shocks should be included in a realistic
modeling of the SZ effect from merging galaxy clusters. Prokhorov et
al. (2010b) incorporated the relativistic Wright formalism for
modeling the SZ effect from a merging galaxy cluster in a numerical
simulation and introduced a method to derive the SZ temperature that
does not require symmetry and so is suitable for merging clusters.

In this paper, we extend the previous studies by introducing a
method to derive the standard temperature deviation and temperature
variance along the line-of-sight for revealing the 3D temperature
structure for relatively cool merging galaxy clusters with
temperatures less than 10 keV. We focus on relatively cool merging
galaxy clusters because it is easier to measure their standard
temperature deviation and temperature variance owing to lower order
SZ corrections.

Shock-heated regions have been found by Chandra in numerous galaxy
clusters in the range from galaxy groups with temperatures of
$\approx$ 1 keV, such a HGC 62 (Gitti et al. 2010), to the most
massive galaxy clusters with temperatures of $\approx$ 15 keV, such
as the Bullet cluster (Markevitch et al. 2002). In this paper, we
show that the temperature variance along the line-of-sight is an
useful quantity, which allows us to reveal shocks in galaxy
clusters, and demonstrate how to reveal a merger shock by analyzing
a relatively cool simulated galaxy cluster.

We also propose an extension of our method to make it suitable for
hot galaxy clusters (in the temperature range of 10 keV -- 15 keV).
We calculate the temperature variance along the line-of-sight for a
simulated hot galaxy cluster to demonstrate how this quantity can be
derived from multi-frequency observations of the SZ effect.

Studying the presence of gas inhomogeneities along the line-of-sight
by means of the SZ effect is important to improve our knowledge of
the 3D temperature structure of merging galaxy clusters. This will
permit us to improve the deprojection analysis of galaxy clusters by
comparing the derived values of the standard temperature deviation
along the line-of-sight with those calculated from the deprojection
maps. An analysis of maps of the gas temperature variance along the
line-of-sight will provide us with an interesting approach to reveal
merger shocks. This approach is alternative to that based on the
projected X-ray surface brightness maps (e.g. Markevitch et al.
2002).

The layout of the paper is as follows. We calculate the SZ intensity
maps of the simulated galaxy cluster at frequencies of 150 GHz and
217 GHz in the framework of the Wright formalism in Sect. 2. We
propose a method to derive the temperature variance along the
line-of-sight for relatively cool galaxy clusters and apply this
method to the simulated galaxy cluster in Sect. 3, using the derived
SZ intensity maps. We also propose a method to reveal merger shocks
in the simulated galaxy cluster in Sect. 3. We present our
discussion on how to improve the proposed method and calculate the
temperature variance for a simulated hot galaxy cluster in Sects. 4
and 5, respectively. We present our conclusions in Sect. 6.

\section{The SZ effect from the simulated cluster}

In this section, we calculate the SZ intensity maps at frequencies
of 150 GHz and 217 GHz in the framework of the relativistic Wright
formalism for the merging galaxy cluster simulated by Dubois et al.
(2010) and previously analyzed by means of the SZ effect by
Prokhorov et al. (2010b). The derived SZ intensity maps will be used
in the next section to calculate the temperature variance along the
line-of-sight.

The galaxy cluster simulations of Dubois et al. (2010), which
we use in this paper, are run with the Adaptive Mesh Refinement
(AMR) code RAMSES (Teyssier 2002). The simulation follows the gas
dynamics using a second-order unsplit Godunov scheme for the Euler
equations with a Total Variation Diminishing scheme for
extrapolation of fluid quantities at the cell interface from their
cell centre values. Particles are evolved with a particle-mesh
solver for gravity and using a Cloud-In-Cell interpolation.

The simulation includes gas cooling from a primordial gas
composition of Hydrogen and Helium with a UV heating background
following Haardt \& Madau (1996). The reionization redshift is set
up at $z=8.5$. Star formation proceeds in high gas-density regions
with $\rho>0.1 \, \rm H\, \rm cm^{-3}$ (Rasera \& Teyssier 2006,
Dubois \& Teyssier 2008). Though feedback from supernovae is not
included, we take into account the feedback from Active Galactic
Nuclei following the model from Dubois et al. (2010), which is the
dominant source of energy in massive objects such as galaxy
clusters.

The simulation is performed assuming a flat $\Lambda$CDM cosmology
(Spergel et al. 2003). A zoom technique is employed with a $128^3$
coarse grid in a 80 $h^{-1}_{100}$ Mpc box size, and with two nested
grids with a 20 $h^{-1}_{100}$ Mpc and a 6 $h^{-1}_{100}$ Mpc radius
for a $256^3$ and $512^3$ equivalent grid respectively, where
$h_{100}=H_{0}/(100\, \rm km\, \rm s^{-1}\, \rm Mpc^{-1})$. This
leads to a minimum dark matter mass of $M_{\rm DM}=4.5\times 10^8\,
\rm M_{\odot}$. The grid is dynamically refined down to level $16$,
reaching 1.19 $h^{-1}_{100}$ kpc. The zoom region tracks the
formation of a galaxy cluster with a 1:1 major merger occurring at z
= 0.8. This z = 0.8 major galaxy merger drives the cluster gas to
temperatures twice the virial temperature thanks to violent shock
waves. The projected mass-weighted temperature map is shown in Fig.
2. For further details of the simulation, see Dubois et al.
(2010).

Using the Wright formalism, we have previously calculated the SZ
intensity maps at frequencies of 128 GHz and 369 GHz for the
simulated cluster at z=0.74 (see Prokhorov et al. 2010b). These
frequencies correspond to minimum and maximum values of the SZ
intensity in the the Kompaneets approximation (Kompaneets 1957). The
ratio of these SZ intensities has allowed us to derive the
SZ-temperature from mock SZ observations. The derived SZ-temperature
is shown in Fig. \ref{Fig1} (top left-hand panel) and demonstrates
how the prominent structures on the 2D projected temperature map of
the merging cluster can be unveiled. The mass-weighted temperature
map (in keV) of the simulated cluster is plotted in Fig. \ref{Fig1}
(top right-hand panel) for a comparison. The gas temperature maps
have prominent ``arc-like'' structures, which have a high
temperature compared with other simulated cluster regions. The
average temperature of the ``arc-like'' structures is $\approx$5 keV
and, therefore, we can use this simulated cluster as an example of a
relatively low temperature galaxy cluster. Below, we briefly
describe how to calculate the SZ intensity maps in the framework of
the Kompaneets and Wright formalisms.

The CMB intensity distortion caused by the SZ effect on a
non-relativistic electron population in the framework of the
Kompaneets approximation is given by (for a review, see Birkinshaw
1999)
\begin{equation}
\Delta I_{\mathrm{nr}}(x) = I_{\mathrm{0}} g(x) y_{\mathrm{gas}},
\label{Inr}
\end{equation}
where $I_{\mathrm{0}}=2 (k_{\mathrm{b}} T_{\mathrm{cmb}})^3 /
(hc)^2$, $x=h\nu/k_{\mathrm{b}} T_{\mathrm{cmb}}$, and the spectral
shape of the SZ distortion is described by the spectral function
\begin{equation}
g(x)=\frac{x^4 \exp(x)}{(\exp(x)-1)^2} \left(x
\frac{\exp(x)+1}{\exp(x)-1}-4\right).
\end{equation}
The subscript  $`\mathrm{nr}' $ denotes that Eq. (\ref{Inr}) was
obtained for a non-relativistic electron population in the
Kompaneets approximation. The Comptonization parameter
$y_{\mathrm{gas}}$ is given by
\begin{equation}
y_{\mathrm{gas}}=\frac{\sigma_{\mathrm{T}}}{m_{\mathrm{e}}c^2} \int
n_{\mathrm{gas}} k T_{\mathrm{e}} dl,
\end{equation}
where the integral must be taken along the line-of-sight,
%extends from the last scattering
%surface of the CMB radiation to the observer at redshift z=0,
$T_{\mathrm{e}}$ is the electron temperature, $n_{\mathrm{gas}}$ is
the number density of the gas, $\sigma_{\mathrm{T}}$ is the Thomson
cross-section, $m_{\mathrm{e}}$ the electron mass, $c$ the speed of
light, $k_{\mathrm{b}}$ the Boltzmann constant, and $h$ the Planck
constant.

The CMB intensity distortion caused by the SZ effect in the
relativistic Wright formalism can be written in the form proposed by
Prokhorov et al. (2010a) given by
\begin{equation}
\Delta I(x) = I_{\mathrm{0}}
\frac{\sigma_{\mathrm{T}}}{m_{\mathrm{e}}c^2} \int n_{\mathrm{gas}}
k_{\mathrm{b}} T_{\mathrm{e}} G(x, T_{\mathrm{e}}) dl, \label{form}
\end{equation}
where $G(x, T_{\mathrm{e}})$ is the generalized spectral function,
which depends explicitly on the electron temperature.

The relativistic spectral function $G(x, T_{\mathrm{e}})$ obtained
in the framework of the Wright formalism is given by
\begin{equation}
G(x, T_{\mathrm{e}})=\int^{\infty}_{-\infty} \frac{P_{1}(s,
T_{\mathrm{e}})}{\Theta(T_{\mathrm{e}})} \left(\frac{x^3 \exp(-3
s)}{\exp(x \exp(-s))-1}-\frac{x^3}{\exp(x)-1}\right) ds, \label{G}
\end{equation}
where $\Theta(T_{\mathrm{e}}) = k_{\mathrm{b}} T_{\mathrm{e}}/
m_{\mathrm{e}}c^2$, and $P_{1}(s, T_{\mathrm{e}})$ is the
distribution of frequency shifts for single scattering (Wright 1979;
Birkinshaw 1999). Note that this form of $G(x, T_{\mathrm{e}})$ is
the most suited for the application to merging clusters because it
uses the local temperature and not an average temperature along the
line-of-sight.

The values of the relativistic $G(x, T_{\mathrm{e}})$ and
non-relativistic $g(x)$ functions are close if the gas temperature
is low $k_{\mathrm{b}} T_{\mathrm{e}}\ll 10$ keV. For relatively
cool galaxy clusters with $k_{\mathrm{b}} T_{\mathrm{e}}\lesssim 10$
keV, the difference between these functions (i.e. the SZ
relativistic correction) is proportional to the parameter
$\Theta=k_{\mathrm{b}} T_{\mathrm{e}}/(m_{\mathrm{e}} c^2)$.
However, as shown in Colafrancesco et al. (2010), the planned SZ
experiments will be able to measure SZ relativistic corrections even
for cool galaxy clusters with temperatures of 3-5 keV.

To calculate the SZ intensity maps in this section, we choose
frequencies of 150 GHz and 217 GHz. The frequency of 150 GHz is
chosen because the SZ effect at this frequency for a relatively cool
cluster with a temperature less than 10 keV is approximately
described by the Kompaneets formalism (we have checked this using
Eqs. (\ref{Inr}) and (\ref{form})). The crossover frequency of 217
GHz at which the SZ effect in the framework of the Kompaneets
approximation equals zero is chosen because it is promising to
analyze SZ relativistic corrections at this frequency. The choice of
frequencies will be explained in more detail in the next section.

\begin{figure}
\centering
\includegraphics[angle=0, width=7.5cm]{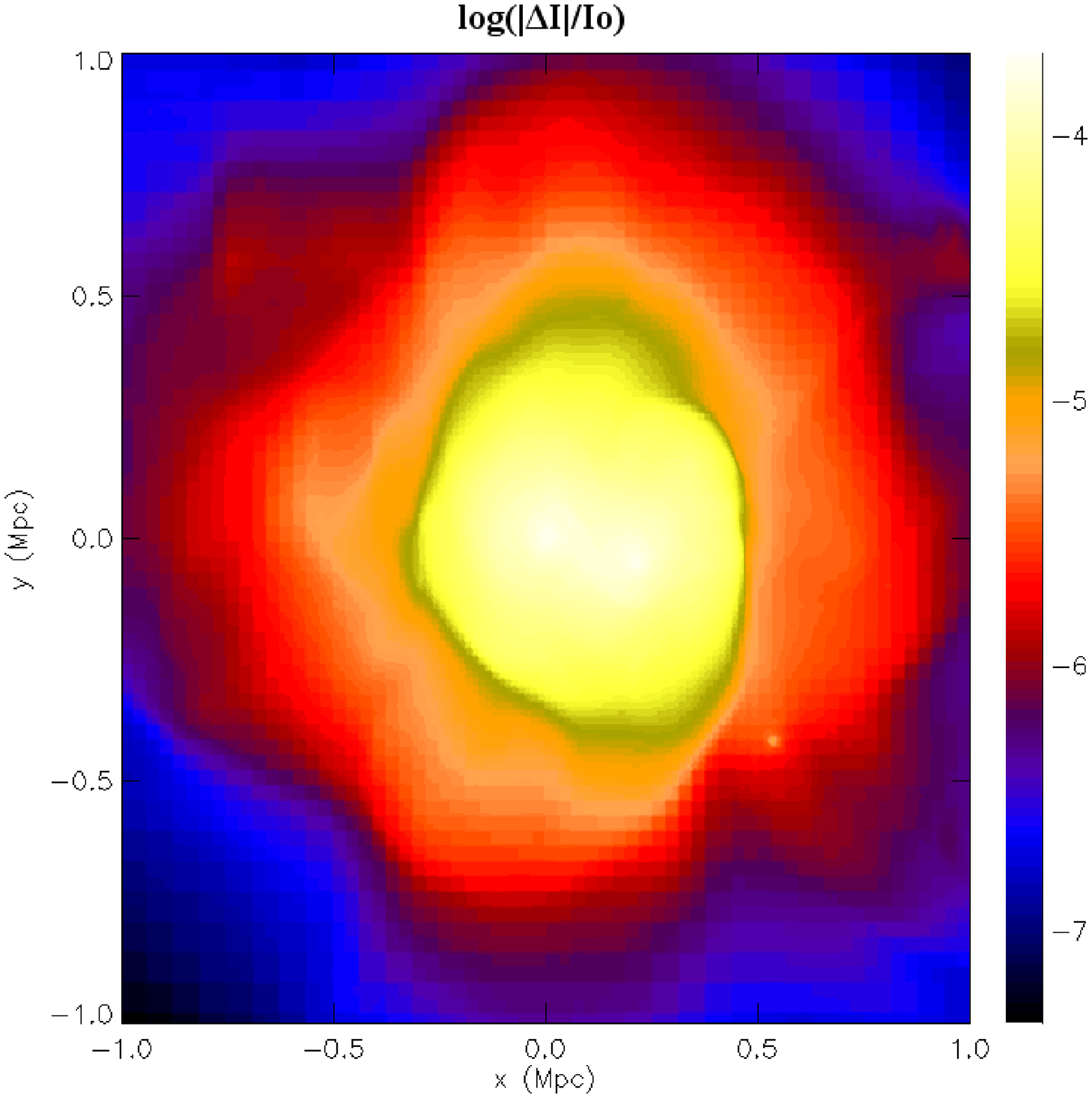}
%\caption{The logarithm of the intensity $|\Delta I|/I_{0}$ of the SZ
%effect at a frequency of 150 GHz derived from the numerical
%simulation in the framework of the Wright formalism.}\label{SZ150}
%\centering
\includegraphics[angle=0, width=7.5cm]{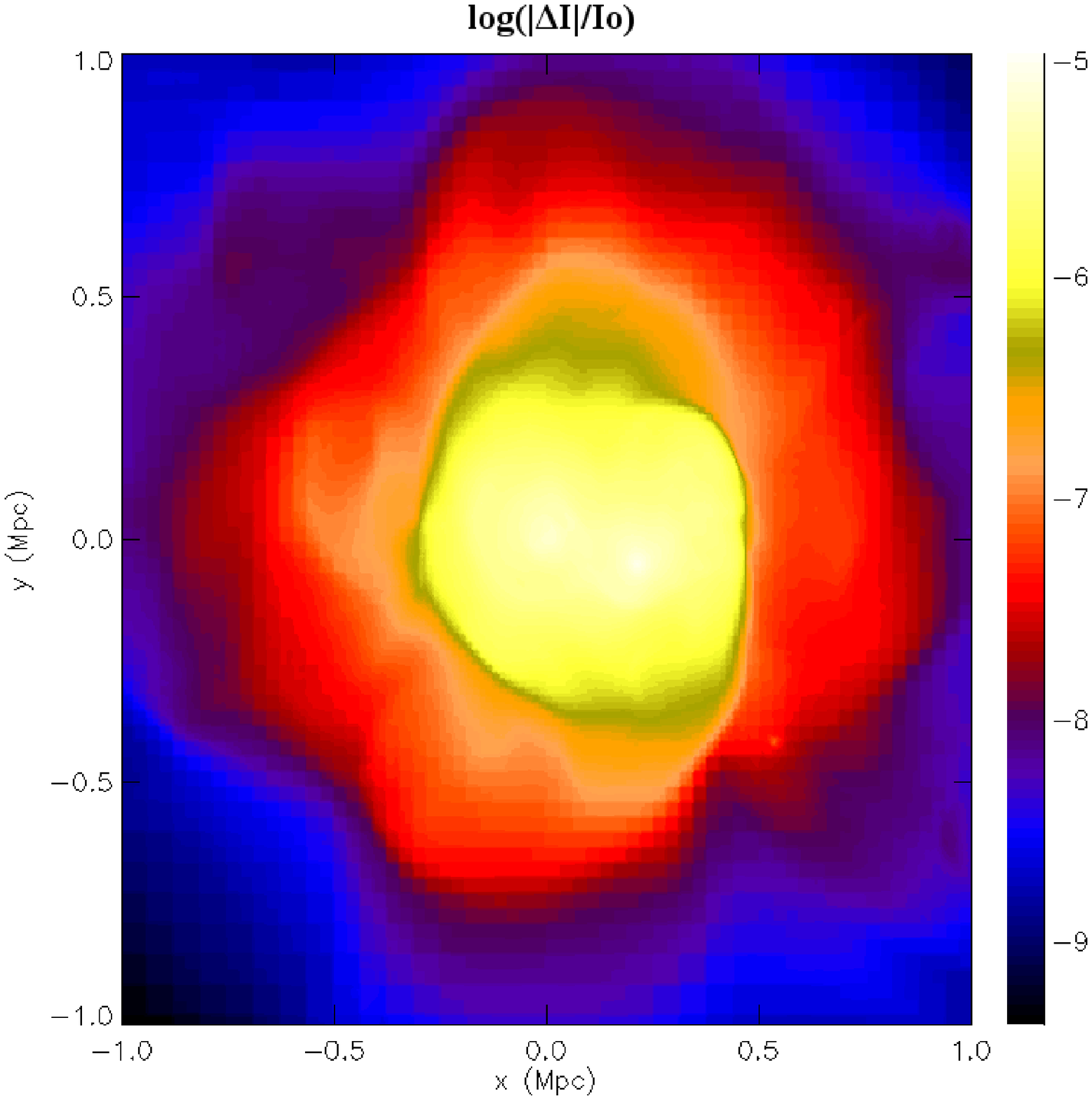}
\caption{The logarithms of the intensity maps $|\Delta I|/I_{0}$ of
the SZ effect at frequency of 150 GHz (upper) and 217 GHz (lower)
derived from the numerical simulation in the framework of the Wright
formalism.}\label{SZ217}
\end{figure}

We use the 3D density and temperature maps for the simulated galaxy
cluster to calculate the SZ effect using the Wright formalism. The
intensity maps of the SZ effect at the frequencies of 150 GHz and
217 GHz derived from the simulation maps of the gas density and
temperature are plotted in Fig. \ref{SZ217}.

In this paper we shall not consider SZ effects due to the motion of
galaxy clusters, we note that the kinematic SZ effect can be
extracted from the SZ intensity map at 217 GHz by using the method
proposed in Sect. 5.1 by Prokhorov et al. (2010b).

%However, note the SZ intensity at a frequency of 217 GHz cannot be
%determined from the Kompaneets approximation and is dominated by SZ
%relativistic corrections, since the SZ signal at 217 GHz in the
%framework of the Kompaneets equation equals zero. Therefore, we have
%calculated the SZ intensity maps at these frequencies.

In the next section, using the derived SZ intensity maps at
frequencies of 150 GHz and 217 GHz, we show how to analyze the 3D
temperature structure of galaxy clusters.

\section{Unveiling the 3D temperature structure of galaxy clusters}

The methods previously used in other papers permit us to derive 2D
temperature maps of galaxy clusters. In this section, we propose a
method to derive the standard temperature deviation and temperature
variance along the line-of-sight for relatively cool galaxy clusters
with temperatures less than 10 keV, which provides us with a tool to
study the 3D temperature structure of merging galaxy clusters by
means of the SZ effect.

Apart from the Wright formalism, there is another formalism to
describe the relativistically correct SZ effect based on an
extension of the Kompaneets equation which allows relativisitic
effects to be included (Challinor \& Lasenby 1998; Itoh et al.
1998). Analytical forms for the spectral changes due to the SZ
effect, correct to first and second order in the expansion parameter
$\Theta=k_{\mathrm{b}} T_{\mathrm{e}}/(m_{\mathrm{e}} c^2)$, are
given in Challinor \& Lasenby (1998). We use the Wright formalism
for calculating the SZ intensity maps from the simulation as
accurately as possible and apply the formalism based on an extension
of the Kompaneets equation to calculate the temperature variance
along the line-of-sight, because it is the only one that can be
manipulated in such a way as to derive this quantity.

We are focusing on studying relatively cool clusters with
temperatures less than 10 keV because the SZ first order correction
to the Kompaneets approximation is sufficient in this temperature
range to calculate the SZ effect with a good precision (i.e. the SZ
second order correction is very small).

The formalism based on an extension of the Kompaneets equation is
obtained by solving the Boltzmann equation using the limitations of
the single scattering approximation. Thus, the Boltzmann equation
(see Challinor \& Lasenby 1998) can only be applied to describe the
SZ effect in optically thin clusters. The single scattering
approximation is usually sufficient to calculate the SZ effect in
galaxy clusters. We have checked that the single scattering
approximation is valid for the considered simulated cluster.

To derive the temperature variance we propose to measure the SZ
effect at two frequencies for disentangling the SZ relativistic
correction from the SZ contribution which can be described by the
Kompaneets approximation. The easiest way to do this is to choose
one frequency at 217 GHz (i.e. the crossover frequency of the SZ
effect in the Kompaneets approximation) and another frequency at
which the SZ effect is described by the Kompaneets approximation.
The SZ relativistic corrections are very small compared with the
total SZ signal for relatively cool clusters at frequencies which
are not close to the crossover frequency and are not very high (we
have checked that the SZ relativistic corrections at frequencies of
$\nu \gtrsim 540$ GHz are higher than 10\% of the total thermal SZ
effect for a galaxy cluster with $k_{\mathrm{b}}T_{\rm e}=5$ keV).
At very high frequencies the contribution of SZ relativistic
corrections is significant. Thus, we choose the second frequency at
150 GHz, which is a good choice since the modern experiments such as
the Atacama Cosmology Telescope
\footnote{http://www.physics.princeton.edu/act/}, Atacama Pathfinder
Experiment telescope \footnote{http://bolo.berkeley.edu/apexsz/},
and South Pole Telescope \footnote{http://pole.uchicago.edu/spt/}
make SZ observations at this frequency.

The CMB intensity change produced by the SZ effect from a galaxy
cluster with inhomogeneous density and temperature distributions in
the formalism based on an extension of the Kompaneets equation,
correct to first order in the expansion parameter $\Theta$, can be
written as

\begin{equation}
\Delta I(x) \approx I_{\mathrm{0}} \sigma_{\mathrm{T}} \int
n_{\mathrm{e}} \Theta\times (g(x)+\Theta\times g_{1}(x)) dl,
\label{form2}
\end{equation}
where $g_{1}(x)$ is the spectral function taken from Eq. (28) of
Challinor \& Lasenby (1998) and is given by

\begin{eqnarray}
g_{1}(x) = \frac{x^4 \exp(x)}{(\exp(x)-1)^2}\left(-10+\frac{47}{2}x
\coth(x/2)-\frac{42}{5}x^2 \coth^2(x/2)+\right.\nonumber\\
\left.\frac{7}{10}x^3 \coth^3(x/2)+\frac{7x^2}{5\sinh^2(x/2)}\left(x
\coth(x/2)-3\right)\right).
\end{eqnarray}

Since the optical depth of the electron plasma $\tau$ can be derived
from X-ray observations or SZ observations, we rewrite Eq.
(\ref{form2}) as

\begin{equation}
\Delta I(x) \approx I_{\mathrm{0}} \tau \left( \frac{\langle
k_{\mathrm{b}} T_{\mathrm{e}}\rangle}{m_{\mathrm{e}} c^2} \times
g(x)+\frac{\langle(k_{\mathrm{b}}T_{\mathrm{e}})^2\rangle}{m^2_{\mathrm{e}}
c^4}\times g_{1}(x)\right), \label{aver}
\end{equation}
where $\langle k_{\mathrm{b}} T_{\mathrm{e}}\rangle=\int
n_{\mathrm{e}}k_{\mathrm{b}}T_{\mathrm{e}} dl/\int n_{\mathrm{e}} dl
$ is the temperature averaged along the line-of-sight and
$\langle(k_{\mathrm{b}} T_{\mathrm{e}})^2\rangle=\int n_{\mathrm{e}}
(k_{\mathrm{b}}T_{\mathrm{e}})^2 dl/\int n_{\mathrm{e}} dl $ is the
squared temperature averaged along the line-of-sight.

As shown by Challinor \& Lasenby (1998), for $k_{\mathrm{b}}
T_{\mathrm{e}}=5$ keV, the second-order effects are very small over
the entire intensity spectrum compared with the total thermal SZ
effect. Therefore, we can use Eq.(\ref{aver}) to calculate
approximately the SZ effect for the simulated cluster.

The SZ intensity at 150 GHz (x=2.64) is approximately given by the
term of Eq.(\ref{aver}) proportional to $\langle k_{\mathrm{b}}
T_{\mathrm{e}}\rangle$, since the SZ relativistic corrections at
this frequency are very small compared with the total thermal SZ
signal, i.e.
\begin{eqnarray*}
\Delta I(2.64) \approx I_{\mathrm{0}} \tau \frac{\langle
k_{\mathrm{b}} T_{\mathrm{e}}\rangle}{m_{\mathrm{e}} c^2} \times
g(2.64).
\end{eqnarray*}
The SZ intensity at 217 GHz (x=3.83) is approximately given by the
term of Eq.(\ref{aver}) proportional to $\langle(k_{\mathrm{b}}
T_{\mathrm{e}})^2\rangle$, because the SZ relativistic corrections
are a dominant contribution at this frequency, i.e.

\begin{eqnarray*}
\Delta I(3.83) \approx I_{\mathrm{0}} \tau
\frac{\langle(k_{\mathrm{b}}T_{\mathrm{e}})^2\rangle}{m^2_{\mathrm{e}}
c^4}\times g_{1}(3.83).
\end{eqnarray*}

%Note that we can use any frequency at which SZ relativistic
%corrections are a small contribution to the SZ effect instead of the
%frequency of 150 GHz.

We note that both the temperature variance along the line-of-sight,
which is given by
\begin{equation}
\mathrm{var}=\langle(k_{\mathrm{b}} T_{\mathrm{e}})^2\rangle-\langle
k_{\mathrm{b}} T_{\mathrm{e}}\rangle^2, \label{varexpr}
\end{equation}
and the standard temperature deviation, which is given by
\begin{equation}
\sigma=\sqrt{\langle(k_{\mathrm{b}} T_{\mathrm{e}})^2\rangle-\langle
k_{\mathrm{b}} T_{\mathrm{e}}\rangle^2},\label{sigmaexpr}
\end{equation}
can be derived for the simulated cluster from SZ observations at
frequencies of 150 GHz (x=2.64) and 217 GHz (x=3.83). The
temperature variance and standard temperature deviation for a
homogeneous gas temperature distribution equal zero and, therefore,
both the temperature variance and standard temperature deviation
along the line-of-sight are a measure of inhomogeneity of the gas
temperature along the line-of-sight.

In terms of the SZ intensities, the temperature variance along the
line-of-sight is given by
\begin{equation}
\mathrm{var}\approx (m_{\mathrm{e}} c^2)^2 \left(\frac{\Delta
I_{\mathrm{217GHz}} }{I_{0}\tau g_{1}(x=3.83)}-\left(\frac{\Delta
I_{\mathrm{150GHz}}}{I_{0}\tau g(x=2.64)}\right)^2\right)
\label{eqvar},
\end{equation}
where $\Delta I_{\mathrm{150GHz}}$ and $\Delta I_{\mathrm{217GHz}}$
are the SZ intensities at frequencies of 150 GHz (x=2.64) and 217
GHz (x=3.83), respectively. The standard temperature deviation can
be derived from SZ intensities at frequencies of 150 GHz and 217 GHz
by using Eq. (\ref{eqvar}) and the relation between the temperature
variance and standard temperature deviation, $\rm var = \sigma^2$.

\begin{figure}
\centering
\includegraphics[angle=0, width=16.0cm]{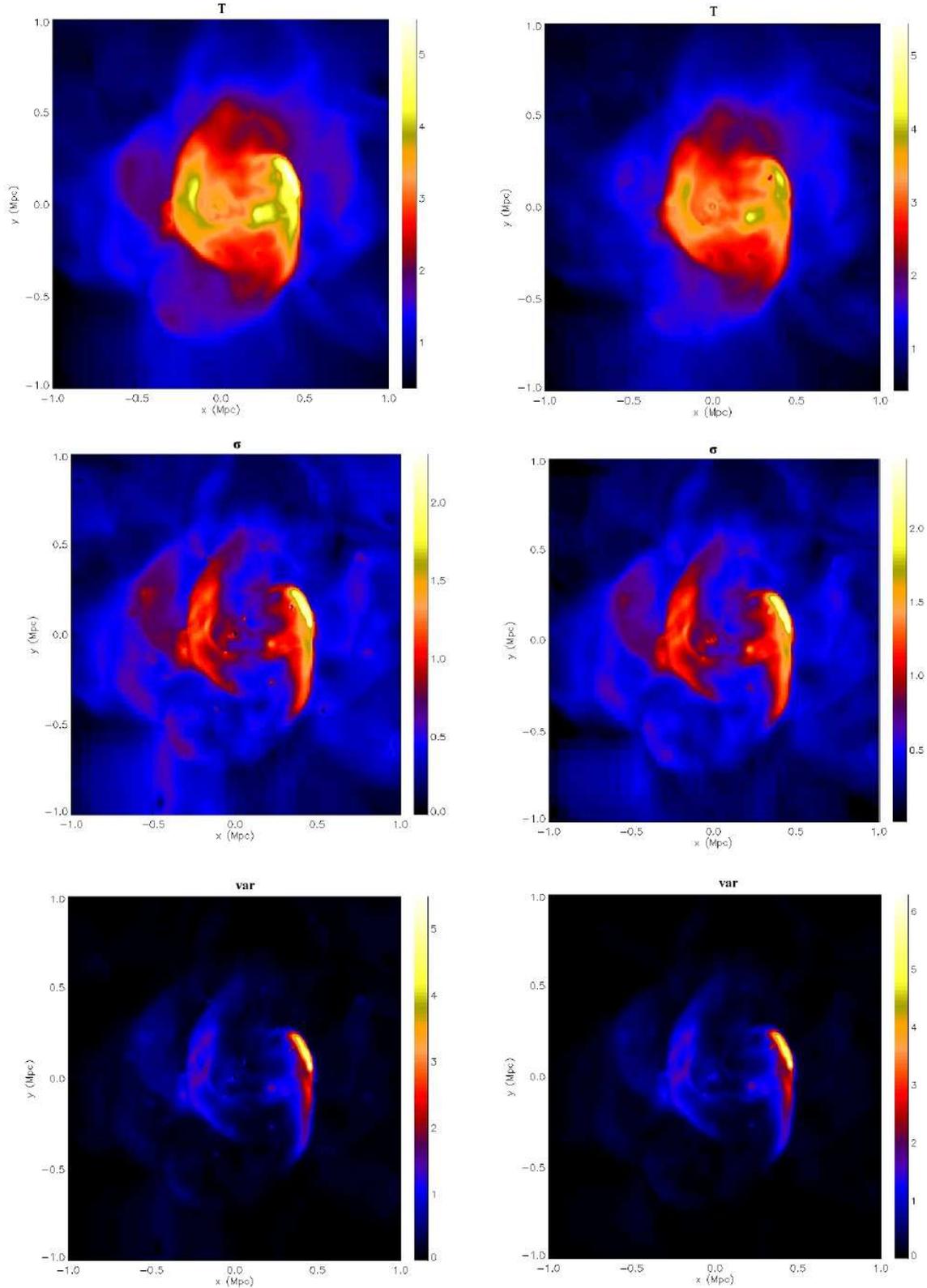}
\caption{Maps of the Compton-average electron temperature in keV
(top left-hand panel), mass-weighted temperature in keV (top
right-hand panel), standard temperature deviation along the
line-of-sight in keV derived from Eq. \ref{eqvar} (middle left-hand
panel), standard temperature deviation along the line-of-sight in
keV derived from 3D gas temperature and density distributions
(middle right-hand panel), temperature variance in keV$^2$ along the
line-of-sight derived from Eq. \ref{eqvar} (bottom left-hand panel),
and temperature variance in keV$^2$ along the line-of-sight derived
from 3D gas temperature and density distributions (bottom right-hand
panel). All the maps are calculated for the simulated relatively
cool galaxy cluster at z=0.74.} \label{Fig1}
\end{figure}

We calculate the map of the standard temperature deviation along the
line-of-sight using Eqs. (\ref{varexpr}), (\ref{sigmaexpr}), and
(\ref{eqvar}) and the SZ intensity maps at frequencies of 150 GHz
and 217 GHz derived in the framework of the Wright formalism in
Sect. 2. The map of the standard temperature deviation along the
line-of-sight is shown in Fig. \ref{Fig1} (middle left-hand panel).

We note that the values of the standard temperature deviation are
high in regions of high temperature on the SZ-temperature map, which
is shown in Fig. \ref{Fig1} (top left-hand panel). The standard
temperature deviation map contains high temperature ``arc-like''
structures, which are also seen in the temperature map. As noted by
Prokhorov et al. (2010b), the origin of these ``arc-like''
structures is the major merger of two galaxy clusters occurring at
z=0.8. Dubois et al. (2010) demonstrates that this major merger
drives the cluster gas to temperatures twice the virial temperature
thanks to violent merger shock waves.

The middle left-hand panel in Fig. \ref{Fig1} shows that the
standard temperature deviation values in the regions of shock waves
are $\simeq$50\% of those of the gas temperature. The increase of
the standard temperature deviation in the ``arc-like'' regions
compared with other regions shows the presence of temperature
inhomogeneities along the line-of-sight in these regions.
\onecolumn

\twocolumn

Our analysis of the standard temperature deviation map shows that
the standard temperature deviation increases in the post-shock
regions. Shock waves are places of significant gas heating,
therefore they are interesting targets for analyzing by means of the
temperature variance which is sensitive to the presence of
temperature inhomogeneities. Using the calculated SZ intensity maps
at frequencies of 150 GHz and 217 GHz and applying Eq.
(\ref{eqvar}), we derive the temperature variance map of the
simulated cluster. The temperature variance map is shown in Fig.
\ref{Fig1} (bottom left-hand panel). Since the temperature variance
equals the squared standard temperature deviation, the regions of
high values of the standard temperature deviation are revealed with
a higher contrast on the temperature variance map than with that on
the standard temperature deviation map. Therefore, an analysis of
the temperature variance map is a way to reveal shock fronts in
merging galaxy clusters.

We conclude that the standard temperature deviation and temperature
variance  are sensitive to the gas temperature distribution along
the line-of-sight and permit us to analyze the 3D temperature
structure, and to reveal merger shock waves in galaxy clusters. We
have demonstrated that the high temperature ``arc-like'' structures
in the temperature maps shown in Fig. \ref{Fig1} (top panels)
correspond to high values of the temperature variance along the
line-of-sight.

\section{Extending the proposed method: higher accuracy and hot clusters}

In the previous section, we have demonstrated how to derive the
standard temperature deviation and temperature variance along the
line-of-sight using the Wright formalism and analytical forms for
the spectral changes due to the SZ effect, correct to first order in
the expansion parameter $\Theta$. In this section, we discuss some
of the limitations of the proposed method, which arise from the use
of analytical forms for the spectral changes correct to first order.
To find the limitations of the proposed method, we calculate the
standard temperature deviation and temperature variance maps using
the simulated 3D gas density and temperature distributions.

The map of the standard temperature deviation along the
line-of-sight, calculated by using the simulated 3D gas temperature
and density distributions instead of the SZ intensities at
frequencies of 150 GHz and 217 GHz, is plotted in Fig. \ref{Fig1}
(middle right-hand panel). Comparing the standard temperature
deviation maps plotted in Fig. \ref{Fig1} (middle panels) shows that
the maximal difference between these standard temperature deviation
maps is around 8\%. Therefore, we conclude that the standard
temperature deviation along the line-of-sight derived from the
method proposed in the previous section is in good agreement with
the true standard temperature deviation along the line-of-sight.

The temperature variance map derived from the 3D gas temperature and
density distributions instead of the SZ intensities is shown in Fig.
\ref{Fig1} (bottom right-hand panel). The maximal difference between
the values of the temperature variance derived from the SZ
intensities (see the bottom left-hand panel in Fig. \ref{Fig1}) and
from the simulated 3D gas temperature and density distributions (see
the bottom right-hand panel in Fig. \ref{Fig1}) is around 15\%. We
propose below how to calculate the variance map by means of the SZ
effect more accurately.

Although the second-order correction is very small compared with the
total SZ effect for relatively cool galaxy clusters, this correction
leads to a $\simeq10\%$ (average) uncertainty on the temperature
variance map. To improve the agreement between the true and derived
temperature variance maps, we propose to use a combined generalized
spectral function proposed in Prokhorov et al. (2010a). Since there
is a small contribution to the SZ signal at a frequency of 217 GHz,
derived in the formalism based on an extension of the Kompaneets
equation, from the second-order effects, we propose to exclude this
contribution by using SZ observations at two frequencies at which
the second-order effects equal to zero. Using Eq. (33) from
Challinor \& Lasenby (2010), we find that the second-order effects
are equal to zero at frequencies of 197 GHz (x=3.47) and 397 GHz
(x=7.0). Therefore, the combined generalized spectral function,
which corresponds to a relativistic contribution to the generalized
function at a frequency of 197 GHz taking into account a measurement
at a frequency 397 GHz and is defined in Eq. (7) of Prokhorov et al.
(2010a), is determined by the first-order effects and does not
depend on the zero-order and second-order effects. If SZ
observations at these frequencies are used to derive the signal of
$\Delta I_{\mathrm{197 GHz}}-\Delta I_{\mathrm{397 GHz}}\times
g(x=3.47)/g(x=7.0)$, which does not depend on the zero-order and
second-order effects, this permits us to use this SZ signal instead
of the SZ signal at 217 GHz to improve the agreement between the
true and derived temperature variance maps. Note that the
third-order corrections are negligible over the entire spectrum for
$k_{\mathrm{b}} T_{\mathrm{e}}\lesssim 10$ keV (Challinor \& Lasenby
1998). We have checked that the contribution of the third-order
corrections to the signal of $\Delta I_{\mathrm{197 GHz}}-\Delta
I_{\mathrm{397 GHz}}\times g(x=3.47)/g(x=7.0)$ is much smaller than
the contribution of the second-order corrections to the SZ signal at
a frequency of 217 GHz. The use of the combined generalized spectral
function permits us to extend the proposed method to hot galaxy
clusters with gas temperatures in the range of 10 keV -- 15 keV (see
Sect. 5).

The method to derive the first-order SZ correction by using SZ
observations at frequencies of 197 GHz and 397 GHz is more
observationally complex than that based on measuring the SZ signal
at the crossover frequency of 217 GHz. This is because the
contamination by point-like sources is lower at 217 GHz than that at
397 GHz and because the 150 GHz and 217 GHz atmospheric windows are
easier for observing from the ground (i.e. the atmosphere is more
transparent at 150 GHz and 217 GHz). Therefore, the method based on
measuring the SZ signal at 217 GHz is more suitable for relatively
cool galaxy clusters with $k_{\mathrm{b}} T_{\mathrm{e}}\lesssim 10$
keV. The advantage of the method based on measuring the SZ signals
at 197 GHz and 397 GHz is that it provides us with a possibility to
measure the temperature variance in hot galaxy clusters with
$k_{\mathrm{b}} T_{\mathrm{e}}$ in the range of 10 keV -- 15 keV.

We conclude that the maps of the standard temperature deviation and
temperature variance derived in Sect. 3 are close to the true maps
derived from 3D gas temperature and density distributions.
Therefore, the maps of the temperature variance and standard
temperature deviation derived from SZ observations at frequencies of
150 GHz and 217 GHz are promising to analyze the 3D temperature
structure of galaxy clusters and to reveal the merger shocks in
galaxy clusters.

An analysis of the standard temperature deviation and temperature
variance along the line-of-sight by means of the SZ effect will
permit us to improve the deprojection technique of merging galaxy
clusters. The deprojection of temperature maps is so far based on
symmetry assumptions, since the quantities integrated along the
line-of-sight are involved. The use of the standard temperature
deviation and temperature variance, which are measure of gas
inhomogeneity along the line-of-sight, provides us with a
possibility to be independent from symmetry assumptions and to
obtain more realistic deprojection temperature maps.

Shock fronts are places of gas heating and, therefore, the gas
temperature inhomogeneities along the line-of-sight are large near
shock fronts. Studying the temperature variance map provides us with
a new independent approach to unveil shock fronts in merging galaxy
clusters.

\section{Studying the 3D temperature structure of hot galaxy clusters}

We now demonstrate how to derive the temperature variance along the
line-of-sight for hot galaxy clusters with gas temperatures in the
range of 10 keV -- 15 keV by using the improved method proposed in
the previous section. To produce the SZ intensity maps at
frequencies of 150 GHz, 197 GHz, and 397 GHz, we use the 3D
numerical simulations of the merging hot galaxy cluster presented in
Akahori \& Yoshikawa (2010).

The simulations of merging galaxy clusters by Akahori \& Yoshikawa
(2010) are carried out using an N-body and SPH code developed in
Akahori \& Yoshikawa (2008), which adopts the entropy--conservative
formulation of SPH by Springel \& Hernquist (2002), and the standard
Monaghan-Gingold artificial viscosity (Monaghan \& Gingold 1983)
with the Balsara limiter (Balsara 1995). They carried out an
adiabatic run in which radiative cooling, star formation, and AGN
feed back are not taken into consideration.

In the simulations, an encounter of two free-falling galaxy
subclusters from the epoch of their turn-around is considered. The
two galaxy subclusters have virial masses and radii of
$M_{\mathrm{200},1}=8\times10^{14} M_\odot$ and
$r_{\mathrm{200},1}=1.91$ Mpc, and
$M_{\mathrm{200},2}=2\times10^{14} M_\odot$ and
$r_{\mathrm{200},2}=1.21$ Mpc, respectively, where $r_{\rm 200}$ is
the radius within which the mean cluster mass density is 200 times
the present cosmic critical density. The initial distributions of a
dark matter halo and a ICM component are the NFW density profile
(Navarro et al. 1997) and the $\beta$-model density profile
(Cavaliere \& Fusco-Femiano 1976), respectively. The scale radius of
the NFW profile is set to $r_{\rm s} = r_{\rm 200}/5.16$, and the
core radius of the $\beta$-model is $r_{\rm c}=r_{\rm s}/2$.

%The radial profiles of velocity dispersion
%of dark matter and temperature of ICM are computed using Jeans
%equation and the assumption of hydrostatic equilibrium,
%respectively. The initial separation of the centers of two galaxy
%subclusters is set to 1.4 times of the sum of their virial radii.
%The corresponding initial relative velocity of the two subclusters
%is calculated using the prescription described in Sarazin (2002).

The massive subcluster is composed of one million dark matter
particles and the same number of SPH particles, while the less
massive subcluster is represented by a quarter million particles for
each component. The corresponding spatial resolution (the smoothing
length) of SPH is $\sim 50$~kpc at the ICM density of $10^{-3} {\rm
cm^{-3}}$. We use the result with zero impact parameter at a time of
$t=0.5$ Gyr, where $t=0$ Gyr corresponds to the time of the closest
approach of the centers of the dark matter halos. The average
temperature of this simulated galaxy cluster is $\simeq$ 10 keV and
the temperature of the hottest regions associated with shock fronts
is $\simeq$ 15 keV. For further details of the simulation, see
Akahori \& Yoshikawa (2010).

The dense cores of these two subclusters are accelerated by the
preceding dark matter halos and shock waves are formed in front of
the cores. Shock fronts are shown in the Mach number distribution
plotted in Fig. 2 of Akahori \& Yoshikawa (2010). Since the
temperature and density maps of the simulated hot merging cluster
are inhomogeneous owing to the presence of shock waves, we study the
map of the temperature variance along the line-of-sight for this
cluster.

For a hot galaxy cluster with inhomogeneous density and temperature
($k_{\rm{b}} T_{\rm{e}}\leq 15$ keV) distributions, the CMB
intensity change produced by the SZ effect in the formalism based on
an extension of the Kompaneets equation, which is correct to second
order in the expansion parameter $\Theta$, can be written as
\begin{equation}
\Delta I(x) \approx I_{\mathrm{0}} \sigma_{\mathrm{T}} \int
n_{\mathrm{e}} \Theta\times (g(x)+\Theta\times
g_{1}(x)+\Theta^2\times g_{2}(x)) dl \label{form22}
\end{equation}
where $g(x)$, $g_{1}(x)$, and $g_{2}(x)$ are the spectral functions
taken from  Challinor \& Lasenby (1998). Note that higher order
terms should be included for calculating the SZ effect from very hot
galaxy clusters with $k_{\rm{b}} T_{\rm{e}}\geq 15$ keV (e.g.
Challinor \& Lasenby 1998; Itoh et al. 1998).
Since the optical depth $\tau$ of the electron plasma can be derived
from X-ray or SZ observations, we rewrite Eq. (\ref{form22}) as
\begin{eqnarray}
&&\Delta I(x) \approx I_{\mathrm{0}} \tau \left( \frac{\langle
k_{\mathrm{b}} T_{\mathrm{e}}\rangle}{m_{\mathrm{e}} c^2} \times
g(x)+\frac{\langle(k_{\mathrm{b}}T_{\mathrm{e}})^2\rangle}{m^2_{\mathrm{e}}
c^4}\times g_{1}(x)+\right. \\
&&\left.\frac{\langle(k_{\mathrm{b}}T_{\mathrm{e}})^3\rangle}{m^3_{\mathrm{e}}
c^6}\times g_{2}(x)\right) \nonumber,\label{aver3}
\end{eqnarray}

\begin{figure}
\centering
\includegraphics[angle=0, width=7.5cm]{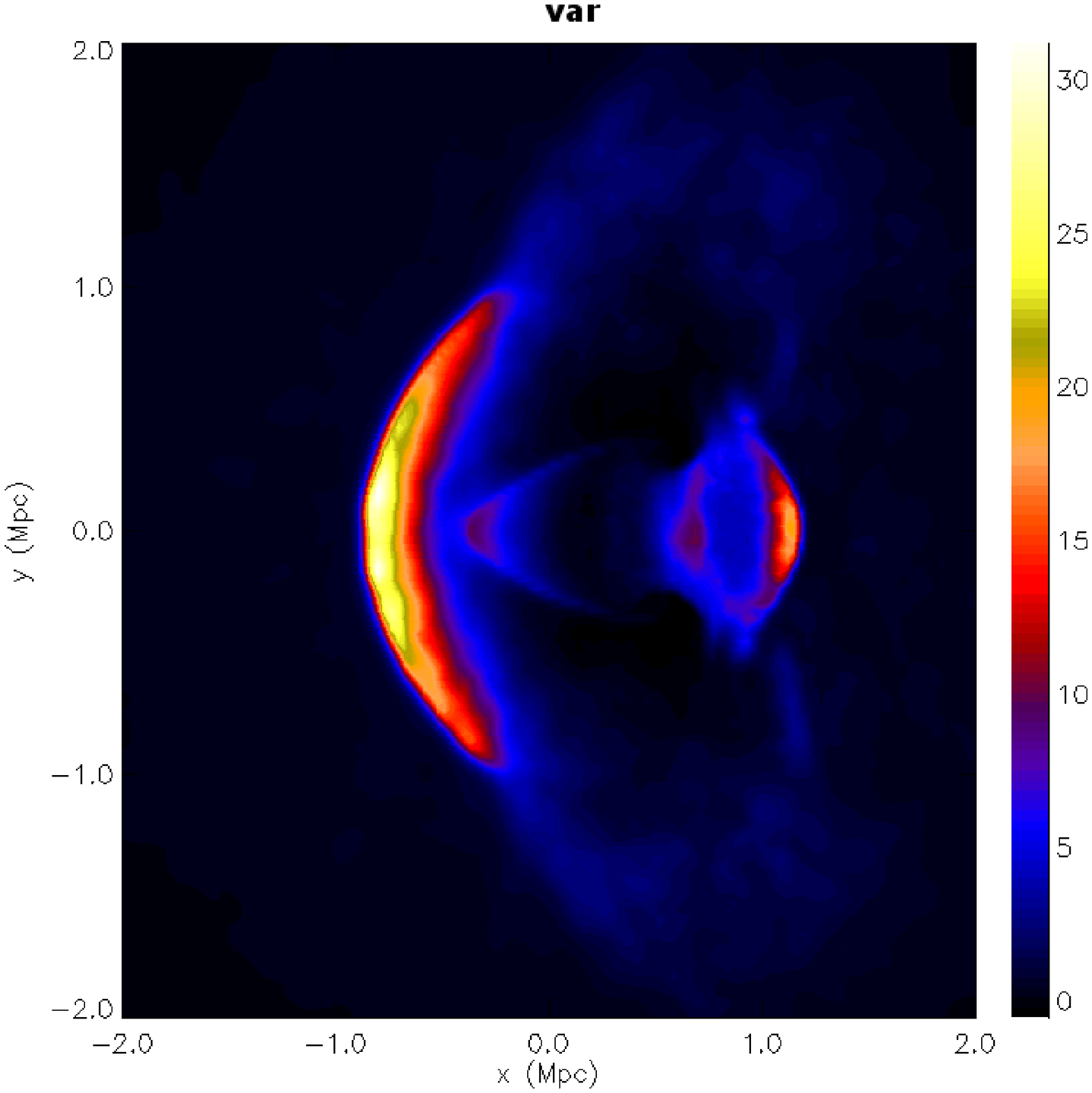}
\includegraphics[angle=0, width=7.5cm]{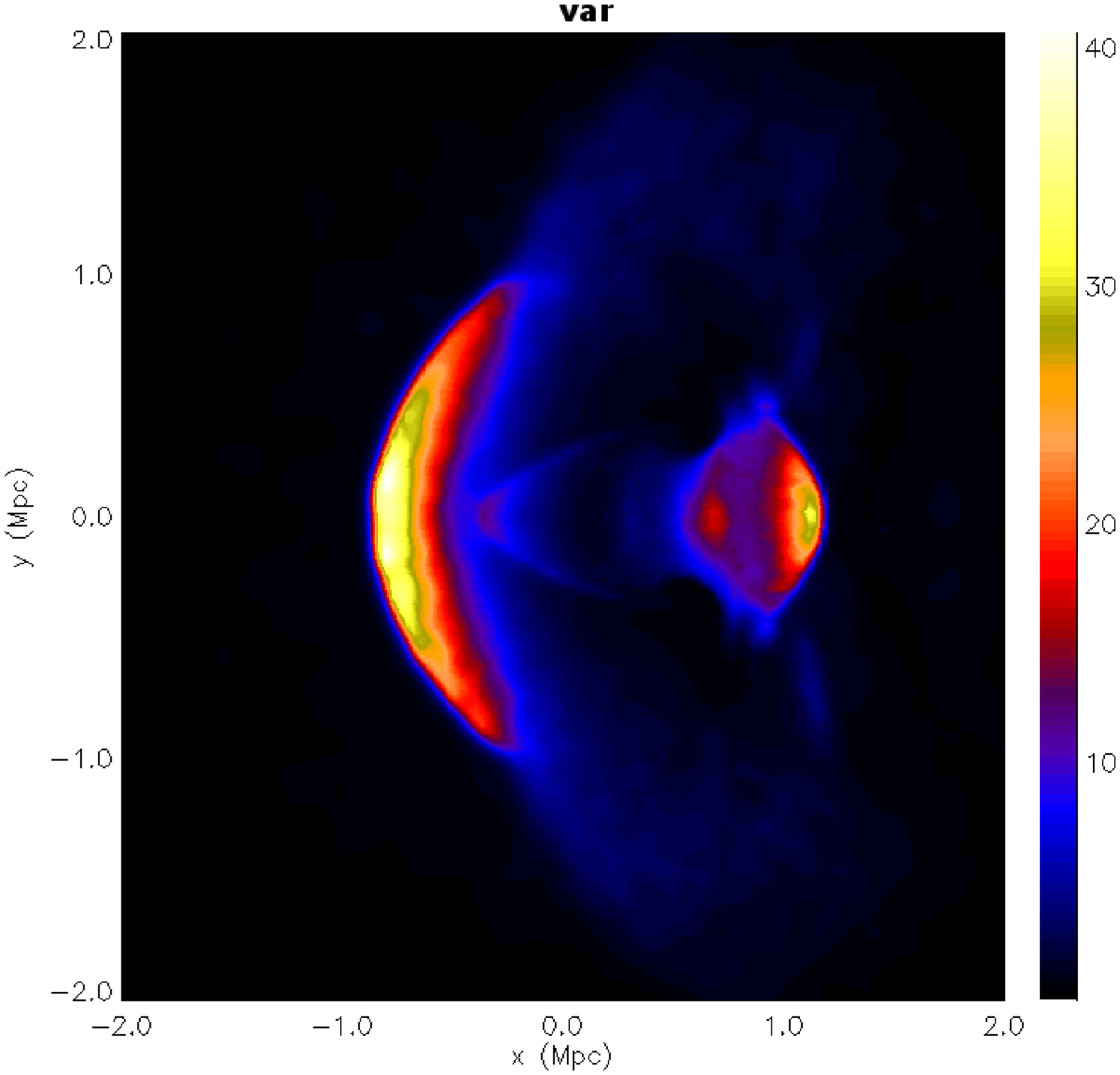}
\caption{Maps of the temperature variance in keV$^2$ along the
line-of-sight derived from Eq. \ref{eqvarhot} (top panel) and from
3D gas temperature and density distributions (bottom
panel)}\label{Fig3}
\end{figure}

In terms of the SZ intensities, the temperature variance along the
line-of-sight is given by
\begin{equation}
\mathrm{var}=(m_{\mathrm{e}} c^2)^2\left(\frac{
\Delta\tilde{I}_{\mathrm{197GHz}, \mathrm{397GHz}} }{I_{0}\tau
\tilde{g}}-\left(\frac{\Delta I_{\mathrm{150GHz}}}{I_{0}\tau
g(x=2.64)}\right)^2\right) \label{eqvarhot},
\end{equation}
where $\Delta I_{\mathrm{150GHz}}$ is the SZ intensity at a
frequency of 150 GHz (x=2.64), $\Delta \tilde{I}_{\mathrm{197GHz},
\mathrm{397GHz}}$ is the combined SZ intensity at frequencies of 197
GHz (x=3.47) and 397 GHz (x=7.0), which is defined as (Prokhorov et
al. 2010a)
\begin{equation}
\Delta\tilde{I}_{\mathrm{197 GHz}, \mathrm{397GHz}}=\Delta
I_{\mathrm{197GHz}}-\Delta I_{\mathrm{397GHz}}\times
\frac{g(x=3.47)}{g(x=7.0)},
\end{equation}
and $\tilde{g}$ is the coefficient which equals
\begin{equation}
\tilde{g}=g_1(x=3.47)-g_{1}(x=7.0)\times\frac{g(x=3.47)}{g(x=7.0)}.
\end{equation}
The map of the standard temperature deviation along the
line-of-sight derived from our mock SZ observations at frequencies
of 150 GHz, 197 GHz, and 397 GHz is shown in Fig. \ref{Fig3} (top
panel).

The map of the temperature variance, calculated by using the
simulated 3D gas temperature and density distributions instead of
the SZ intensities at frequencies of 150 GHz, 197 GHz, and 397 GHz,
is plotted in Fig. \ref{Fig3} (bottom panel). Comparing the maps in
Fig. \ref{Fig3} shows that the maximal difference between these maps
is $\approx$ 25\%. Measurements of the temperature variance by means
of the proposed method permit us to study temperature
inhomogeneities along the line-of-sight. We find that the values of
the the temperature variance are high in the post-shock regions and,
therefore, the map of this quantity can also be used to unveil
merger shocks in hot galaxy clusters.

We will study possibilities of deriving the temperature variance
along the line-of-sight for hot merging galaxy clusters by means of
the planned SZ experiments in a forthcoming paper.

\section{Conclusions}

A number of important gasdynamic processes can be studied by
observing merging galaxy clusters. Studying the temperature
structure of merging clusters is important to reveal heated regions
associated with merger shock waves and is complicated because the
temperature maps of merging clusters are very inhomogeneous and
because the approximation of a spherically symmetric gas
distribution is not suitable for merging galaxy clusters. In many
merging clusters, such as the Abell 3376 cluster (see Bagchi et al.
2007), the temperature distribution is very inhomogeneous and it is
possible to derive only the 2D temperature map.

In this paper we demonstrate how to derive the quantities describing
the temperature distribution along the line-of-sight for relatively
cool merging galaxy clusters with $k_{\mathrm{b}}
T_{\mathrm{e}}\lesssim 10$ keV. To produce a realistic merging
cluster, we have used a zoomed cosmological simulation from Dubois
et al. (2010) and found that the simulated galaxy cluster undergoes
a violent merger. The temperature map of the simulated cluster
becomes very inhomogeneous because of the merger activity. We have
incorporated the relativistic Wright formalism for modeling the SZ
effect in the numerical simulation using the algorithm proposed by
Prokhorov et al. (2010a). We calculated the SZ intensity maps at
frequencies of 150 GHz and 217 GHz.

We propose a method to derive the standard temperature deviation and
temperature variance along the line-of-sight using SZ observations
at frequencies of 150 GHz and 217 GHz. We found that the values of
the standard temperature deviation in the regions of shock waves are
around 50\% of those of the gas temperature. The increase of the
values of the standard temperature deviation in the ``arc-like''
regions, which were studied by Prokhorov et al. (2010b), compared
with other regions shows the presence of temperature inhomogeneity
along the line-of-sight in the ``arc-like'' regions. Using the 3D
gas temperature and density distributions in the simulated galaxy
cluster, we have checked that the map of the standard temperature
deviation derived by means of the proposed method based on SZ
observations is in agreement with the true map of the standard
temperature deviation.

We show that the temperature variance along the line-of-sight is an
useful quantity which provides us with a possibility to reveal
merger shock waves in galaxy clusters. We calculated this quantity
by using SZ observations at frequencies of 150 GHz and 217 GHz. The
``arc-like'' structures are revealed with a much higher contrast in
the map of this quantity than that in the map of the standard
temperature deviation and shock fronts can be observed by measuring
this quantity. Therefore, an analysis of maps of the temperature
variance is interesting to reveal merger shock waves in galaxy
clusters.

We extend our method and make it applicable to hot merging galaxy
clusters with gas temperatures in the range of 10 keV -- 15 keV in
Sect. 4. Using the data of numerical simulations of Akahori \&
Yoshikawa (2010), we demonstrate how to derive the temperature
variance along the line-of-sight for the simulated hot merging
cluster of galaxies in Sect. 5.

Our study demonstrates that the SZ effect is a promising tool to
reveal the 3D temperature structure of galaxy clusters. We have
proposed the method which provides us with the possibility of an
analysis of the temperature distribution along the line-of-sight.

 %\citet{b9}.

\section*{Acknowledgments}
We are grateful to Sergio Colafrancesco and Neelima Sehgal for
discussions and thank the referee for valuable suggestions.

\bsp

\label{lastpage}

\end{document}